\documentclass[aps,prb,twocolumn,groupedaddress]{revtex4}

\usepackage{graphicx}
\usepackage{bm}

\begin{document}



\title{Magnitude and crystalline anisotropy of hole magnetization in (Ga,Mn)As}



\author{C.\ \'Sliwa}
\email{sliwa@ifpan.edu.pl}
\affiliation{Institute of Physics, Polish Academy of Sciences, al.\ 
  Lotnik\'ow 32/46, PL 02-668 Warszawa, Poland}

\author{T.\ Dietl}
\email{dietl@ifpan.edu.pl}
\affiliation{Institute of Physics, Polish Academy of Sciences and ERATO 
Semiconductor Spintronics Project, Japan Science and Technology, 
al.\ Lotnik\'ow 32/46, PL 02-668 Warszawa, Poland}
\affiliation{Institute of Theoretical Physics, Warsaw University, 
PL 00-681 Warszawa, Poland}


\date{\today}

\begin{abstract}
  Theory of hole magnetization $M_\mathrm{c}$ in zinc-blende
  diluted ferromagnetic semiconductors is developed 
  relaxing the spherical approximation of earlier 
  approaches. The theory is employed to determine $M_\mathrm{c}$ for (Ga,Mn)As over
  a wide range of hole concentrations and a number of crystallographic orientations of 
  $\mathrm{Mn}$ magnetization. It is found that anisotropy of $M_\mathrm{c}$
  is practically negligible but the obtained magnitude of $M_\mathrm{c}$
  is significantly greater than that determined in the spherical approximation. Its
  sign and value compares favorably with the results of available magnetization measurements and 
  ferromagnetic resonance studies.
\end{abstract}

\pacs{}

\maketitle

\section{Introduction}

Following the discovery of ferromagnetism in zinc-blende Mn-based
semiconducting compounds,\cite{Dietl2003} a theory has been developed 
that describes correctly a number of
properties of those materials.\cite{Dietl2001,Jungwirth2006b} The theory, 
which is based on the Zener
model of ferromagnetism mediated by band carriers, takes into account the complex structure
of the valence band of zinc-blende semiconductors {\em via} the
$\bm{k}\cdot\bm{p}$ method, and employs the molecular-field
and virtual-crystal approximations to include the exchange interaction
between the Mn \textit{d} electrons and the valence band holes.\cite{Dietl2001,Dietl2000} In
particular, the magnetic moment of the hole carriers per unit volume,
$M_c(p)$, has been calculated for a range of hole
concentrations~$p$.\cite{Dietl2001} The results,\cite{Dietl2001} 
corroborated recently by an independent calculation,\cite{Jungwirth2006a} 
have shown that while a part of this magnetic moment
can be attributed to spin polarization of the hole Fermi liquid, the
orbital magnetic moment of the holes has also to be taken into
account. However, in order to reduce a differential multi-band Landau equation to 
an algebraic one,  it has been assumed 
in Refs.~\onlinecite{Dietl2001} and~\onlinecite{Jungwirth2006a}
that two relevant Luttinger parameters have the same value, $\gamma_2
=\gamma_3$. A quantitative error introduced but this approximation
has been hard to estimate and, moreover, within such an approximation a significant warping
of the valence band has been neglected. 

Motivated by the recent accurate magnetization measurements
carried out by Sawicki and co-workers\cite{Jungwirth2006a}
and ferromagnetic resonance studies of Liu {\em et al.},\cite{Liu2005} both 
completed for high quality (Ga,Mn)As, we have decided to 
develop theory of hole magnetization in carrier-controlled ferromagnetic
semiconductors valid for the arbitrary values of $\gamma_2$ and
$\gamma_3$. Our results show that in the experimentally relevant
range of hole concentrations the theoretical values of $M_c(p)$
are much greater that those evaluated previously,\cite{Dietl2001,Jungwirth2006a}
which allows us for a better description of the experimental findings.\cite{Jungwirth2006a,Liu2005}
Furthermore, we find that the crystalline anisotropy of $M_c(p)$ is practically
negligible, so that the dipole interaction between the subsystems of Mn
and hole magnetic moments does not contribute to magnetic anisotropy
of (Ga,Mn)As.

The calculation proceeds by the determination
of the partition function of the holes occupying the Landau levels in
in the presence of Mn spontaneous magnetization $\bm{M}$, assumed to be parallel 
to the external magnetic field~$\bm{B}$.  
Hole magnetization is then obtained by differentiating the Gibbs thermodynamic potential with
respect to~$\bm{B}$. In contrast to the case $\gamma_2
=\gamma_3$, a significant amount of symbolic calculations involving
trigonometric expressions is required, followed by an efficient
numerical solution of a truncated infinite eigenvalue problem. This
methodology allows us to calculate and examine hole magnetization for
various crystallographic directions of the magnetic field and
magnetization of the $\mathrm{Mn}$-sublattice.

The remaining part of this paper is organized as follows: in Section~II we
present the details of the valence band structure of a zinc-blende
semiconductor; in Section~III we discuss quantized motion of an
electron in uniform magnetic field; in Section~IV we describe some
technicalities of our algorithm; in Section~V we present our results,
and finally in Section~VI we present conclusions of our work.

\section{The $\bm{k}\cdot\bm{p}$ model of the valence band}

\begin{table*}[htb]
  \centering
  \begin{tabular}{@{\extracolsep{3em}}ccc}
    $T_x = \frac{1}{3\sqrt{2}} \pmatrix{-\sqrt{3}& 0& 1& 0\\ 0& -1& 0&
    \sqrt{3}}$& $T_y = \frac{-i}{3\sqrt{2}} \pmatrix{\sqrt{3}& 0& 1&
    0\\ 0& 1& 0& \sqrt{3}}$& $T_z = \frac{\sqrt{2}}{3} \pmatrix{0& 1&
    0& 0\\ 0& 0& 1& 0}$\\
    $T_{xx} = \frac{1}{3\sqrt{2}} \pmatrix{0& -1& 0& \sqrt{3}\\
      -\sqrt{3}& 0& 1& 0}$& $T_{yy} = \frac{1}{3\sqrt{2}} \pmatrix{0&
      -1& 0& -\sqrt{3}\\ \sqrt{3}& 0& 1& 0}$& $T_{zz} =
    \frac{\sqrt{2}}{3} \pmatrix{0& 1& 0& 0\\ 0& 0& -1& 0}$\\
    $T_{yz} = \frac{i}{2\sqrt{6}} \pmatrix{-1& 0& -\sqrt{3}& 0\\ 0&
      \sqrt{3}& 0& 1}$& $T_{zx} = \frac{1}{2\sqrt{6}} \pmatrix{-1& 0&
      \sqrt{3}& 0\\ 0& \sqrt{3}& 0& -1}$& $T_{xy} = \frac{i}{\sqrt{6}}
    \pmatrix{0& 0& 0& -1\\ -1& 0& 0& 0}$
  \end{tabular}
  \caption{Matrices for the cross space of the valence and split-off
    valence band states ($U_i = T_i^{\dagger}$, $U_{ij} =
    T_{ij}^{\dagger}$).}
  \label{tab1}
\end{table*}

Following classic works of Luttinger\cite{Luttinger1955} as well as of
Bell and Rogers\cite{Bell1966} and Pidgeon and Brown,\cite{Pidgeon1966}
the energy states in semiconductors of zinc-blende structure in a magnetic field
(specifically $\mathrm{InSb}$) have been considered in detail by
Trebin, R\"ossler, and Ranvaud,\cite{Trebin1979} who obtained
the effective mass Hamiltonian taking into account the two-fold
degenerate conduction band of the
symmetry~$\Gamma_6$, the four-fold degenerate uppermost valence band
of the symmetry~$\Gamma_8$, and the two-fold degenerate split-off
valence band of the~$\Gamma_7$ symmetry. A model in which an upper $\Gamma_8$
and $\Gamma_7$ conduction bands are explicitly included was developed 
by Pfeffer and Zawadzki.\cite{Zawadzki1996}

Here, we consider explicitly only the
valence band, namely the bands of the $\Gamma_8$ and~$\Gamma_7$
symmetry, denoted respectively as $v$ and~$s$, for which we choose the
basis $\left(\left| \frac32, \frac 32 \right>, \left| \frac32, \frac 12
\right>, \left| \frac32, -\frac 12 \right>, \left| \frac32, -\frac 32
\right>, \left| \frac12, \frac 12 \right>, \left| \frac12, -\frac 12
\right>\right)$.\footnote{Our basis is related to that of
  Ref.~\onlinecite{Dietl2001} Appendix~A as follows: $u_1 = -\left|
    \frac32, \frac 32 \right>$, $u_2 = -i\left| \frac32, \frac 12
  \right>$, $u_3 = \left| \frac32, -\frac 12 \right>$, $u_4 = i\left|
    \frac32, -\frac 32 \right>$, $u_5 = \left| \frac12, \frac 12
  \right>$, $u_6 = -i\left| \frac12, -\frac 12 \right>$.} Therefore,
in our $\bm{k}\cdot\bm{p}$ model of the valence band of a zinc-blende
semiconductor, the band electron Hamiltonian has the block form
\begin{equation}
  \mathcal{H} = \pmatrix{\mathcal{H}^{vv}& \mathcal{H}^{vs}\cr
    \mathcal{H}^{sv}& \mathcal{H}^{ss}},
\end{equation}
where
\begin{eqnarray}
  \mathcal{H}^{vv} & = & -\frac{\hbar^2}{m} \Big\{ \frac12 \gamma_1 k^2 - \gamma_2[
  (J_x^2 - \frac13 J^2) k_x^2 + c.p.] \nonumber \\
  & & {} -2 \gamma_3[ \{J_x, J_y\} \{k_x, k_y\} + c.p.] \Big\} \nonumber \\
  & & {} - \frac{e\hbar}{m} \kappa ( \bm{J} \cdot \bm{B} ), \\
  \mathcal{H}^{ss} & = & -(\Delta_0 + \frac{\hbar^2}{2m} \gamma_1 k^2)
  \nonumber \\
  & & {} - \frac{e\hbar}{m} ( \kappa + \frac{g_0}{4} ) (
  \bm{\sigma} \cdot \bm{B} ), \\
  \mathcal{H}^{vs} & = & \frac{\hbar^2}{m}[ -3\gamma_2(U_{xx} k_x^2 +
  c.p.) \nonumber \\
  & & {} - 6\gamma_3(U_{xy} \{k_x, k_y\} + c.p. )] \nonumber \\
  & & {} + \frac32 \frac{e\hbar}{m}(\kappa+\frac{g_0}{2}) (
   \bm{U} \cdot \bm{B} ).
\end{eqnarray}
Here, $\{ A, B \} = \frac12 (AB + BA)$ and c.p.\ denotes cyclic
permutations. The matrices $J_i$ are the spin-$\frac32$ angular
momentum matrices, $\sigma_i$ are the spin-$\frac12$ angular momentum
matrices multiplied by~2 (Pauli matrices in our basis), and $T_i$,
$T_{ij}$ are the cross space matrices introduced in
Ref.~\onlinecite{Trebin1979}, in our basis given in Table~\ref{tab1}
(which corrects one misprinted sign in Ref.~\onlinecite{Trebin1979},
Table~~I). Notice the Zeeman terms of the form given in
Ref.~\onlinecite{Dietl2001} that properly accounts for the Land\'e
factor of the free electron, $g_0$, a correction in the overall sign
of $\mathcal{H}^{vs}$, as well as the factor of two in the first
(kinetic) term of $\mathcal{H}^{vs}$ that has changed with respect to
Ref.~\onlinecite{Trebin1979}.

Now, to complete our model with the exchange interaction between the
valence band carriers and the localized $d$-electron spins, treated
within the molecular-field and virtual-crystal approximations,
we augment our Hamiltonian with the $p$-$d$ exchange matrix
$\mathcal{H}_{pd}$,
\begin{equation}
  \mathcal{H}_{pd} = B_G \pmatrix{ 2 ( \bm{J} \cdot \bm{w} )&
  6 ( \bm{U} \cdot \bm{w} )\cr
  6 ( \bm{T} \cdot \bm{w} )&
  - ( \bm{\sigma} \cdot \bm{w} )},
\end{equation}
where the $B_G$ parameter is proportional to the $\mathrm{Mn}$
magnetization~$M$ and the $p$-$d$ coupling constant~$\beta$,
\begin{equation}
  B_G = A_F\frac{\beta M}{6 g \mu_B},
\end{equation}
where $A_F \approx 1.2$ is the Fermi liquid Landau parameter 
describing effects of the hole-hole exchange interaction,\cite{Dietl2001,Jungwirth2006b}
and $\bm{w}$ is the versor pointing in the direction of the
$\mathrm{Mn}$~magnetization; see Ref.~\onlinecite{Dietl2001},
Eqs.~A13 to A18.

\section{Quantized motion of an electron in uniform magnetic field}

The differential operators $k_i$ in our $\bm{k}\cdot\bm{p}$ model are
\begin{equation}
  k_\alpha = -i \, \frac{\partial}{\partial x_\alpha} + \frac{e A_\alpha}{\hbar},
\end{equation}
and satisfy $\bm{k} \times \bm{k} = e \bm{B} / i \hbar$, or
\begin{equation}
  \varepsilon_{\alpha\beta\gamma} k_\beta k_\gamma = \frac{e B_\alpha}{i \hbar}.
\end{equation}
Following Luttinger\cite{Luttinger1955}, by
\begin{eqnarray}
  a & = & \frac{1}{\sqrt{2s}}(k_1 - i k_2), \\
  a^{\dagger} & = & \frac{1}{\sqrt{2s}}(k_1 + i k_2),
\end{eqnarray}
we introduce a pair of operators $(a, a^\dagger)$ satisfying the
canonical commutation relation, $[a, a^{\dagger}] = 1$. Here, $s = e B
/ \hbar$, and the coordinate system ``1'', ``2'', ``3'' is such that
$\bm{B}$ is along the ``3'' direction. Then, $k_H = k_3$ (the momentum
component along the direction of the magnetic field) commutes with $a$
and $a^{\dagger}$. We will express the effective mass Hamiltonian in
terms of $(a, a^{\dagger}, k_H)$.

In the Landau gauge, the eigenvalues of $n = a^{\dagger} a$, together
with $k_1$ and~$k_3$, number the energy levels (Landau levels) for a
free electron in the uniform magnetic field~$\bm{B}$. The
corresponding wavefunctions are given
by\cite{LandauLifshitz1958}
\begin{equation}
  \psi = \exp[i (k_1 x_1 + k_3 x_3)] u_n(x_2 - \bar x_2),
\end{equation}
where $u_n(y) = (s/\pi)^{1/4} \exp(-s y^2/2) H_n(\sqrt{s} y) / \sqrt{2^n n!}$ 
is the wavefunction of
a harmonic oscillator, and $\bar x_2 = k_1/s$. Since the vector
potential $\bm{A}(\bm{x}) = (-B x_2, 0, 0)$ does not depend on
$x_1$ and~$x_3$, we use periodic boundary conditions for those two
variables. As far as $x_2$ is concerned, the particle is localized by
the magnetic field, therefore the boundary conditions can be ignored
(at least in the thermodynamic limit, {\em i.e}.\ when the magnetic length
is much smaller than the size of the system). On the other hand, the
length of the system in the ``2'' direction determines the range of
integration over~$k_1$. Hence, we have the formula (B3) of
Ref.~\onlinecite{Dietl2001}:
\begin{eqnarray}
  G_c & = & - k_B T \frac{e B}{2 \pi \hbar} 
   \int_{-\infty}^{\infty} \frac{dk_H}{2\pi} \nonumber \\
   & & \sum_{i=0}^\infty
   \log\{1+\exp[\frac{\varepsilon_i(k_H)-\varepsilon_F}{k_B T}]\},
  \label{eqnGc}
\end{eqnarray}
where we have changed the sign under exponent to reflect the fact that
our particles are holes, while $\varepsilon_i(k_H)$ is the $i$-th
energy level for an electron with $k_3 = k_H$ (the result differs by
an additive constant corresponding to the valence band fully occupied
by electrons).

\section{Details of the algorithm}

Calculation of $G_c$ according to Eq.~(\ref{eqnGc}) requires
precise knowledge of the energy levels~$\varepsilon_i(k_H)$. In
contrast to the case $\gamma_2 = \gamma_3$, the eigenproblem
for energy levels does not decompose into $6\times6$ blocks. Therefore, we
have to generate a truncated matrix of the Hamiltonian, and select
from its eigenvalues only those that approximate eigenvalues of the
infinite matrix. Our Hamiltonian is of the form $\mathcal{H} = h_1
\otimes 1 + h_n \otimes n + h_a \otimes a + {h_{a}}^{\dagger} \otimes
{a}^{\dagger} + h_{a^2} \otimes a^2 + {h_{a^2}}^{\dagger} \otimes
{{a}^{\dagger}}^2$, where $h_1$, $h_n$, $h_a$, and $h_{a^2}$ are $6\times6$
complex matrices with entries being trigonometric expressions in the
angles $(\theta, \phi)$ specifying the direction of the magnetic
field.

\begin{figure}[tb]
  \centering
  \includegraphics[width=0.95\columnwidth]{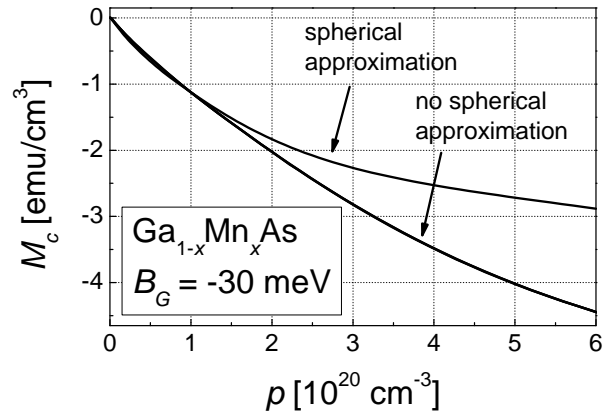}
  \caption{Hole liquid contribution to volume magnetization,
    $\bm{M}_c(p)$ as a function of the hole density $p$
    for $\gamma_2 = \gamma_3$ (``spherical approximation'',
    Refs.~\onlinecite{Dietl2001}, \onlinecite{Jungwirth2006a})
    and for the real band structure parameters of $\mathrm{GaAs}$
    (``no spherical approximation'').
    The spin splitting corresponds to the saturation value
    of magnetization for Mn concentration $x = 0.05$. The calculations
    have been carried out for Mn magnetization along
    three principal crystallographic directions.}
  \label{fig1}
\end{figure}

\begin{figure}[tb]
  \centering
  \includegraphics[width=0.95\columnwidth]{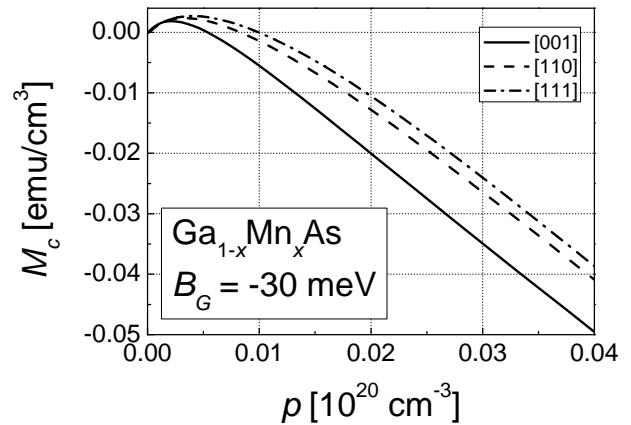}
  \caption{The curves of Fig.~\ref{fig1} at low hole densities
    assuming various directions of  $\mathrm{Mn}$ magnetization.}
  \label{fig2}
\end{figure}

Instead of just truncating the operators $(a, a^{\dagger})$ at
some~$n_{max}$, we employ a procedure that is accurate for $\gamma_2 =
\gamma_3$ in the sense that the generated matrix of the Hamiltonian is
truncated at the non-zero $6\times6$-block boundary. This requires
applying to $\mathcal{H}$ a unitary transformation $\mathcal{H} \to
\mathcal{R} \mathcal{H} \mathcal{R}^{-1}$, where $\mathcal{R} =
\mathcal{R}(\theta, \phi)$ is a $6\times6$ unitary matrix that
implements a rotation from the crystal back to the ``1'', ``2'', ``3''
coordinate system, so that in the spherical case the resulting matrix
does not depend on $(\theta, \phi)$. Then, since in the spherical case
the wavefunctions assume the form $(c_1 u_{n}, c_2 u_{n+1}, c_3
u_{n+2}, c_4 u_{n+3}, c_5 u_{n+1}, c_6 u_{n+2})$, where $c_i$ are the
unknown components of the eigenvectors, from the transformed
$6(n_{max}+1) \times 6(n_{max}+1)$ matrix we drop columns and rows
numbered $-18$, $-12$, $-11$, $-8$, $-6$, $-5$, $-4$, $-2$, $-1$ (the
sign minus means
counting from the right/bottom), so as to preserve a number of whole
$6\times6$ blocks in which (in the spherical case) the Hamiltonian
matrix is non-zero.

The generated matrix is diagonalized using the LAPACK algorithm for
band Hermitian matrices (only the eigenvalues are
computed)\footnote{See Ref.~\onlinecite{Lang1993} for an alternative,
parallelizable, algorithm that has been implemented for real
symmetric matrices and probably can be generalized to complex
Hermitian matrices.}.  An eigenvalue is then selected as correct if
for two subsequent sizes of the truncated matrix we obtain eigenvalues
that are the same within the numerical precision (this is called the
``second truncation'' test in Ref.~\onlinecite{Bell1966}). Great care
must be undertaken because the set of selected eigenvalues does not
have to be complete unless $n_{max}$ is large enough, {\em i.e}.\ there may
be some lacking eigenvalues in the range between the smallest and the
largest eigenvalues selected. The minimal selected eigenvalue does not
monotonically decrease with increasing~$n_{max}$ in such a case.  That
is why we choose $n_{max} \approx 5000$ as a starting point {\em e.g}.\ for
the [110] direction of the magnetization~$\bm{M}$.

\section{Results}

We have performed computations for (Ga,Mn)As adopting the 
previously employed values of the band structure parameters and 
p-d exchange integral.\cite{Dietl2001}
Figure~\ref{fig1} presents the determined values of the hole
magnetization for three principal crystallographic directions 
of magnetization, compared to the data obtained for the
case $\gamma_2 = \gamma_3=2.58$, which reproduce the earlier results.\cite{Dietl2001,Jungwirth2006a} 
The value of the spin splitting parameter $B_G = -30\, \mathrm{meV}$ has been assumed, which corresponds 
to an effective Mn concentration $x_{\mathrm{eff}} = 0.05$ if magnetization
is saturated.
As seen, the curves for the three crystallographic directions [001], [110]
and~[111]  of  $\mathrm{Mn}$ magnetization
overlap and are indistinguishable.  This demonstrates that a strong warping
of the valence band does not result in anisotropy of the hole magnetization
in the experimentally important range of the hole concentrations $p$.
A magnified view of the region near $p = 0$ is presented
in Fig.~\ref{fig2} to show that in this range a small anisotropy 
of $\bm{M}_c = 0$ becomes visible.

\begin{figure}[b]
  \centering
  \includegraphics[width=0.95\columnwidth]{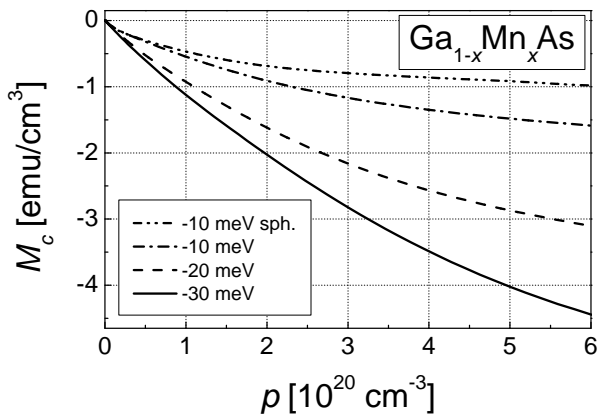}
  \caption{Hole magnetization $\bm{M}_c(p)$ for various magnitudes of valence
  band exchange splitting ($B_G$ equal $-10$, $-20$, and~$-30$ $\mathrm{meV}$).}
  \label{fig3}
\end{figure}

The dependence of the carrier magnetization $\bm{M}_c$ on the
$\mathrm{Mn}$ magnetization~$\bm{M}$ is presented in
Fig.~\ref{fig3}. Three curves for different values of the
parameter $B_G$ ($-10$, $-20$, and~$-30$ $\mathrm{meV}$) are displayed.
For comparison, the data for $B_G =10\, \mathrm{meV}$ and $\gamma_2 = \gamma_3=2.58$
are also shown. An important aspect of the results depicted in Figs.~\ref{fig1} and \ref{fig3}
is that the approximation $\gamma_2 = \gamma_3=2.58$ employed previously underestimates
significantly the magnitude of the hole magnetization. An accurate comparison
between experimental and theoretical results is, however, somewhat hampered by a relatively large uncertainty
in the experimental value of $M_c$, whose determination from magnetization measurements requires 
accurate information on the effective Mn concentration. Nevertheless, as an example we
consider two samples with $x = 0.05$, {\em i.e}.\ $B_G = -30 \,
\mathrm{meV}$, and hole concentrations $p = 4.4 \cdot 10^{20}$ and $8.4 \cdot
10^{20}$ $\mathrm{cm}^{-3}$. For these samples the previous model predicts
$M_c =-0.25$ and  $-0.32$ $\mu_B$ per one Mn ion, respectively. 
The current theory leads to $M_c =-0.36$ and $-0.49$~$\mu_B$.
These sets of results can be compared to the experimental data, $M_c =-0.36$ and $-0.80$ $\mu_B$,  
which we have obtained by taking
a mean value of $M_c$, as measured for $x = 0.045$ and $0.056$
before and after annealing.\cite{Jungwirth2006a} We see that the present
theory describes better the experimental findings but, as mentioned above, more
quantitative comparisons requires detail information on the concentration
of Mn spins that are not compensated by the antiferromagnetic coupling
to interstitial Mn neighbors.

\begin{table}[tb]
  \centering
  \begin{tabular}{@{\extracolsep{6pt}}cccc}
    \hline
    Sample No.& 1& 2& 3\\
    \hline
    \multicolumn{4}{c}{Experiment}\\
    $n_h$-MF [$\mathrm{cm}^{-3}$]& $1.24 \cdot 10^{20}$& $1.48 \cdot
    10^{20}$& $1.64 \cdot 10^{20}$\\
    $g_{\mathrm{eff}}$& $1.92 \pm 0.04$& $1.87 \pm 0.03$& $1.80 \pm
    0.02$\\
    \hline
    \multicolumn{4}{c}{Theory}\\
    $\mathcal{P}$& 0.836& 0.828& 0.822\\
    $M_c/\mu_B$ [$\mathrm{cm}^{-3}$]& $-1.46 \cdot 10^{20}$& $-1.69
    \cdot 10^{20}$& $-1.84 \cdot 10^{20}$\\
    $g_{\mathrm{eff}}$& 1.90& 1.89& 1.88\\
    \hline
  \end{tabular}
  \caption{Comparison of effective Land\'e factor determined
  by ferromagnetic resonance (Ref.~\onlinecite{Liu2005}) to theoretical
  values obtained in the present work.}
  \label{tab2}
\end{table}

Another relevant experiment is ferromagnetic resonance, which leads
to the magnitude of the effective Land\'e factor systematically smaller than 2,
an effect taken as evidence for the hole contribution to the magnetization
dynamics.\cite{Liu2005}
A comparison of our calculations with the experimental data of
Ref.~\onlinecite{Liu2005} is given in Table~\ref{tab2}. The
theoretical values of~$g_{\mathrm{eff}}$ have been calculated assuming
$n_{\mathrm{Mn}} = 1.01 \cdot 10^{21} \, \mathrm{cm}^{-3}$, $B_G = -30
\, \mathrm{meV}$, and using the formula similar to Eq.~2 of
Ref.~\onlinecite{Liu2005}:
\begin{equation}
  \frac{Sg_{\mathrm{Mn}}n_{\mathrm{Mn}}  + M_c/\mu_{B}}{g_{\mathrm{eff}}} =
  Sn_{\mathrm{Mn}} + sn_h \mathcal{P},
\end{equation}
where   $S = 5/2$, $g_{\mathrm{Mn}} = 2.0$, $s = 1/2$,
and $\mathcal{P}$ is the hole liquid spin
polarization defined in Eq.~11
of Ref.~\onlinecite{Dietl2001}. We see that our theory describe satisfactorily 
the deviation of the value of the effective Land\'e factor from 2,
although for the sample with the highest~$T_{\mathrm{C}}$
the deficit of the magnetization is still larger than our calculation
predicts. This disagreement may result from an underestimation of the experimental
value of the hole concentration that was \emph{calculated} based on the value of~$T_{\mathrm{C}}$.\cite{Liu2005}

\section{Conclusions}

In our work, presumably for the first time, an efficient
approach has been proposed allowing for  the calculation of the Landau level energies in
the valence band within the $\bm{k}\cdot\bm{p}$ model of
zinc-blende materials for an arbitrary direction of the magnetic field
and non-zero hole momentum $k_{\mathrm{H}}$ taking effects of cubic anisotropy into account.

The model has been employed to evaluate the contribution of
kinetic and spin energies 
to the volume magnetization of the holes in (Ga,Mn)As. Surprisingly, 
the evaluated crystalline anisotropy
of the carrier magnetization in the relevant range of the hole densities is negligibly small.
At the same time we have found that the 
magnitude of $M_c$ is about 50\% larger than that calculated earlier in the
spherical approximation. We have demonstrated that our improved theory describes
better the available results of magnetization measurements\cite{Jungwirth2006a} and magnetic resonance\cite{Liu2005}
studies.

\begin{acknowledgments}
This work was supported in part by ERATO Semiconductor Spintronics Project
of Japan Science and Technology Agency and NANOSPIN E.~C. project (FP6-2002-IST-015728). 
The computations has been in part performed at the
Interdisciplinary Centre for Mathematical and Computational
Modelling, Pawinskiego 5A, PL-02-106, Warsaw.
\end{acknowledgments}

\end{document}